\documentclass[journal=aamick, manuscript=article]{achemso}
\usepackage{graphicx}
\usepackage{caption}
\usepackage[version=3]{mhchem}
\usepackage[T1]{fontenc}
\usepackage{textgreek}
\usepackage{amsmath}
\usepackage{stackengine}
\usepackage[symbol]{footmisc}
\usepackage[labelfont=bf]{caption}
\usepackage[section]{placeins}
\usepackage[utf8]{inputenc}
\usepackage[export]{adjustbox}
\usepackage{comment}
\usepackage{subfig}
\usepackage[]{hyperref}
\usepackage{cleveref}
\usepackage{amsmath}
\usepackage{mwe}
\usepackage{amsfonts}
\usepackage{ragged2e}
\usepackage{float}
\usepackage{textcomp}
\usepackage{xcolor}
\usepackage[british]{babel}
\usepackage{csquotes}
\DeclareUnicodeCharacter{0308}{ }
\author{Ahsan Noor}
\affiliation{Center for Biomolecular Nanotechnologies, Istituto Italiano di Tecnologia, Via Barsanti 14, 73010 Arnesano (LE), Italy}
\altaffiliation
{Dipartimento di Ingegneria Elettrica e dell\' Informazione, Politecnico di Bari, Via Re David 200, 70125 Bari, Italy}
\author{Anoop R. Damodaran}
\email{rdanoop@umn.edu}
\author{In-Ho Lee}
\affiliation
{Department of Electrical and Computer Engineering, University of Minnesota, Minneapolis, Minnesota 55455, United States}
\author{Stefan A. Maier}
\affiliation{
Chair in Hybrid Nanosystems, Nanoinstitut Munich, Faculty of Physics, Ludwig-Maximilians Universit\"{a}t M\"{u}nchen, K\"{o}niginstrasse 10, 80539 M\"{u}nchen}
\altaffiliation{
Experimental Solid State Physics Group, Department of Physics, Imperial College London, London SW7 2AZ}
\author{Sang-Hyun Oh}
\affiliation
{Department of Electrical and Computer Engineering, University of Minnesota, Minneapolis, Minnesota 55455, United States}
\author{Cristian Cirac\`i}
\email{cristian.ciraci@iit.it}
\affiliation{Center for Biomolecular Nanotechnologies, Istituto Italiano di Tecnologia, Via Barsanti 14, 73010 Arnesano (LE), Italy}

\title{Mode-matching enhancement of second-harmonic generation with plasmonic nanopatch antennas}
\abbreviations{SH, SHG, FP1, FP2, FP3}

\begin{document}
\pagebreak
\begin{abstract}
Plasmonic enhancement of nonlinear optical processes confront severe limitations arising from the strong dispersion of metal susceptibilities and small interaction volumes that hamper desirable phase-matching-like conditions. Maximizing nonlinear interactions in nanoscale systems require simultaneous excitation of resonant modes that spatially and constructively overlap at all wavelengths involved in the process.
Here, we present a hybrid rectangular patch antenna design for optimal second harmonic generation (SHG) that is characterized by a non-centrosymmetric dielectric/ferroelectric material at the plasmonic hot spot.
The optimization of the rectangular patch allows for the independent tuning of various modes of resonances that can be used to enhance the SHG process.
We explore the angular dependence of SHG in these hybrid structures and highlight conditions necessary for maximal SHG efficiency.
Furthermore, we propose a novel configuration with a periodically-poled ferroelectric layer for orders-of-magnitude enhanced SHG at normal incidence. Such a platform may enable the development of integrated nanoscale light sources and on-chip frequency converters.
\end{abstract}

\pagebreak
Optical functionalities achieved through the nonlinear interaction of light with matter are cornerstones of many present-day technological innovations\cite{garmire2013nonlinear}. 
These include control over the laser spectrum (optical frequency conversion), ultra-short pulse generation, and all-optical signal processing\cite{boyd2008nonlinear}.
Optical nonlinear susceptibilities of natural materials are intrinsically low, and conventional nonlinear optical devices rely on high laser intensities and long propagation distances in macroscopic crystals in order to exhibit sizable nonlinear effects\cite{boyd2008nonlinear}.
Such devices are often not compatible with an integrated design, and consequently hinder the realization of efficient nanoscale nonlinear optical components, which are essential for all-optical signal processing in photonic integrated circuits.
To this end, resonant excitation (electromagnetic field enhancements) of nonlinear dielectric based nanoresonators \cite{Carletti15,liu2016resonantly,Gili:2016hl,sautter2019tailoring,Koshelev288}, plasmonic metamaterials \cite{bouhelier2003near,klein2006second,canfield2007local,zeng2009,ko2011nonlinear,harutyunyan2012enhancing,Kolkowski14,argyropoulos2014enhanced,lassiter2014third,huang2015optical,kravtsov2016plasmonic,dass2019gap,deng2020giant}, and hybrid metal-dielectric metamaterials/metasurfaces and waveguides \cite{fan2006second,barakat2012theoretical,aouani2014third,lehr2015enhancing,hentschel2016linear,linnenbank2016second,shibanuma2017efficient,nielsen2017giant,wang2017efficient,wang2019enhancing,qin2019enhanced,shen2020active} have been proposed to improve the efficiency of nonlinear optical processes in small volumes.

Among nonlinear optical processes, achieving efficient frequency conversion at the nanoscale is particularly desirable for many applications in biosensing \cite{deka2017nonlinear}, photonic circuitry \cite{nielsen2017giant}, and quantum optics \cite{caspani2017integrated}.
The difficulty in realizing efficient frequency conversion at the nanoscale arises from the fact that some of the factors that contribute to the wave-mixing processes are often hard to satisfy simultaneously.
In particular, a nanosystem needs to fulfill three main requirements \cite{celebrano2015mode,wang2017efficient}: i) generate local field enhancement, through the excitation of resonant modes, at all the wavelengths involved in the nonlinear process \cite{thyagarajan2012enhanced,aouani2012multiresonant,celebrano2015mode,de2019difference}; ii) the different modes at the frequencies of interest need to exhibit significant spatial overlap in order to maximize their interaction in the nonlinear volume\cite{wang2017efficient,wang2019enhancing}; and iii) nonlinear polarization currents need to constructively add up and efficiently couple to the far-field \cite{butet2015optical}.
A nanosystem that fulfills all the aforementioned properties and simultaneously offers the possibility of realizing an experimentally viable design, may pave the way towards the realization of the efficient nanoscale nonlinear devices.

Plasmonic structures can be used for nonlinear optics in two distinct configurations\cite{Kauranen:2012ff}: i) a pure nonlinear-plasmonic configuration, in which the intrinsic nonlinear responses of the metals in the system are exploited \cite{klein2006second,canfield2007local,zeng2009,czaplicki2013enhancement,celebrano2015mode,Wells:2018fl,czaplicki2018less,scalora2018harmonic,de2019difference}, and ii) a hybrid plasmonic-dielectric configuration, where plasmonic enhancement is used to enhance the nonlinear responses of optically active dielectric materials\cite{argyropoulos2014enhanced,huang2015optical,lassiter2014third,lehr2015enhancing,fan2006second}.
Although metals may posses large second- and third-order nonlinear susceptibilities, their opaqueness makes the design of pure nonlinear-plasmonic configurations challenging.
Moreover, second-order nonlinear processes require a break of symmetry both at the microscopic level (i.e., at the metal surface) and at the macroscopic level (structure asymmetry) to avoid near- and far-field cancellation.
This condition is difficult to achieve in some of the most efficient plasmonic systems, often characterized by locally symmetric gaps.
This includes the case of two nanoparticles a few nanometers apart (in a dimer configuration) that are known to demonstrate some of the largest local field enhancements \cite{PhysRevLett.52.1041,aubry2011plasmonic}.
Likewise, plasmonic systems formed by metallic nanoparticles over a metallic film, such as film-coupled nanosphere \cite{leveque2006optical,mock2008distance,ciraci2012probing} and nanopatch antennas \cite{moreau2012controlled,hao2010high} retain the characteristics of a dimer configuration in terms of local field enhancements, while simultaneously offering a more precise control over the thickness of gaps using modern fabrication techniques such as layer-by-layer deposition \cite{mock2008distance, ciraci2012probing} and atomic layer deposition \cite{ciraci2014film}.
Compared to its film-coupled nanosphere counterpart, plasmonic film-coupled nanopatch systems posses a richer mode structure.
Film-coupled nanopatch systems support \textit{gap-plasmon} modes that are induced between the flat face of the nanopatch antenna and the metallic film \cite{lassiter2013plasmonic,dass2019gap}.
The unique properties of these modes include a wider range of tunability of resonances through careful selection of various design parameters of the system (e.g., the size of the nanopatch or the gap between the film and patch), and efficient far-field coupling due to the magnetic-dipole-like emission pattern of the patch antenna system \cite{PhysRevB.93.081405,rose2014control,ciraci2013quasi}.
Encompassing robust resonant response \cite{lassiter2013plasmonic}, efficient free-space coupling \cite{rose2014control,ciraci2013quasi}, and relative ease of fabrication and incorporation of optically active dielectric gap materials, the nanopatch antenna system is an ideal candidate for developing efficient on-chip nonlinear devices.

To bolster the efficiency of the nonlinear processes involving plasmonic components, often a single resonance is matched either with the fundamental wavelength to enhance the pump intensities \cite{klein2006second, zeng2009,qin2019enhanced} or with the generated harmonic wavelength \cite{metzger2015strong} to enhance the emission efficiency.
This approach has been employed to demonstrate enhanced third-order nonlinear optical processes in film-coupled nanopatch antennas \cite{argyropoulos2014enhanced,huang2015optical} and its two-dimensional counterpart, i.e., film-coupled nanowires \cite{lassiter2014third,liu2015clarifying,qin2019enhanced}, with plasmonic resonances tuned at the pump wavelengths.
Other nanoantenna designs have been proposed to realize doubly- or multi-resonant designs for SHG, sum- and difference-frequency generation \cite{celebrano2015mode,de2019difference}.
Recently, nanopatch antennas have been exploited to demonstrate enhanced SHG through coupling between the gap-plasmon mode of the nanopatch system and epsilon-near-zero mode of the spacer layer \cite{dass2019gap}, and simultaneous control of third-harmonic generation, sum-frequency generation, and four-wave mixing\cite{shen2020active}.

In this article, we present a numerical investigation of mode-matched SHG from plasmonic nanopatch antennas.
The proposed system operates in the hybrid framework, with a thin dielectric spacing layer of a non-centrosymmetric material acting as source of nonlinearity within a plasmonic structure.
First, characteristics of the linear response of the plasmonic system and the modes taking part in mode-matched SHG are discussed.
Linear resonant characteristics of two optimized mode-matched configurations of distinct modal interactions and their SHG efficiency spectra are then introduced and analyzed.
We show how the symmetry of the modes taking part in the nonlinear process might lead to higher or lower SHG efficiency and associate this behaviour to the maximization/minimization of the \textit{overlap integral}, a key parameter in nonlinear emission process.
Finally, we present an ideal system that maximize SHG efficiency through optimization of this integral.

\begin{figure}
\centering
\includegraphics[width=0.8\linewidth]{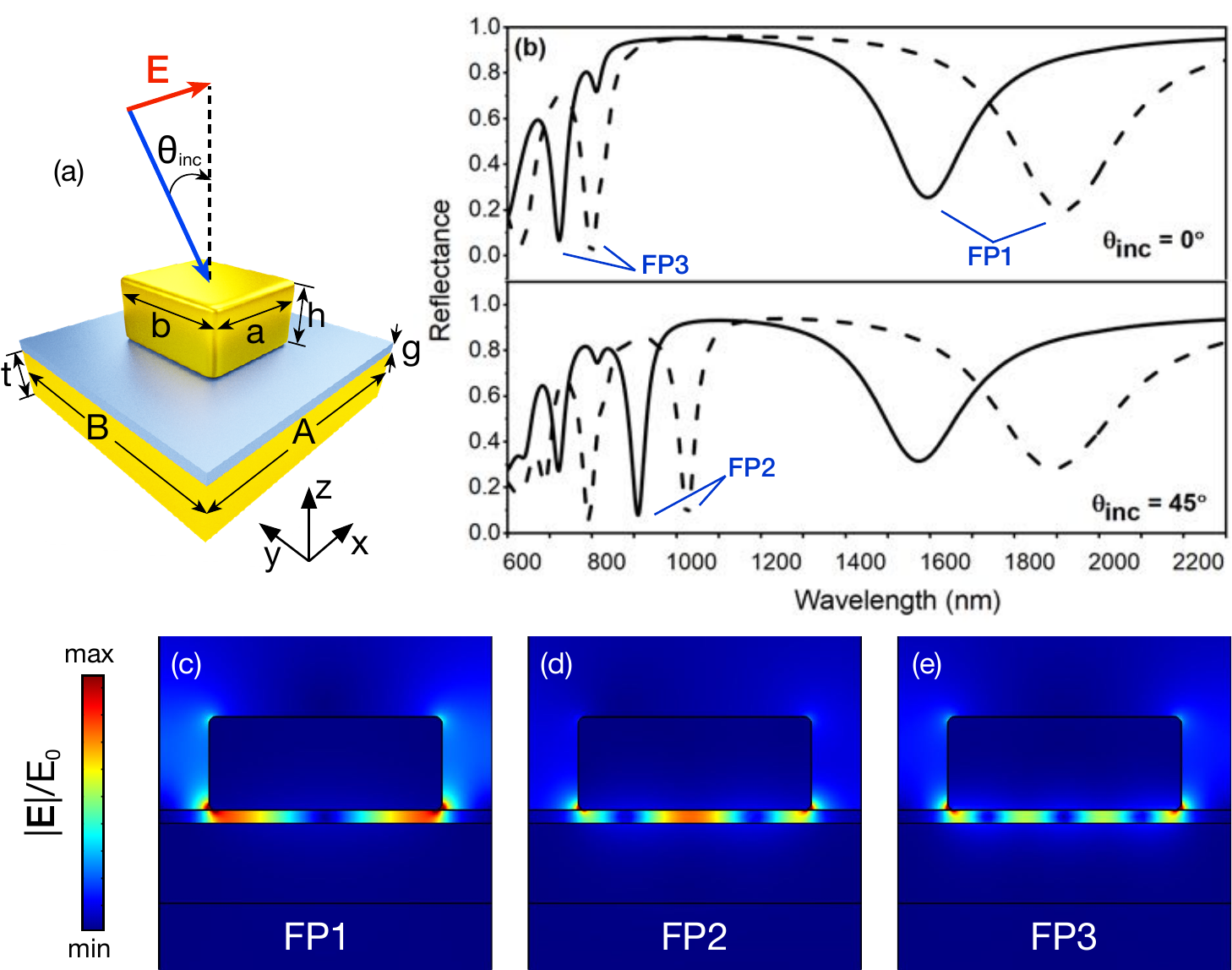}
\caption{Device layout and its linear electromagnetic response. (a) Schematic of the unit-cell of  film-coupled nanopatch system and illustration of its design parameters. (b) Simulated linear reflectance spectra for different values of the patch width $a$ ($b$ is unchanged), and (c-e) the normalized electric field distributions of the modes indicated as FP1, FP2, and FP3 in (b).}
\label{fig:1}
\end{figure}

As already stated, the nanopatch antenna system offers a variety of resonant modes that can be used to enhance nonlinear interactions.
The wavelengths associated to these modes can be tuned by acting on the geometrical parameters of the system.
In particular, we consider a periodic array whose unit-cell design consist of a rectangular gold patch coupled to a gold substrate through a dielectric layer, as illustrated in \Cref{fig:1}a.
We consider a HfO$_2$-based ferroelectric spacer that can be grown using mature ALD processes with excellent CMOS compatibility and potential for on-chip integration \cite{dicken2008electrooptic,muller2011ferroelectricity}.

We perform numerical calculations using a commercial software based on the finite-element method and incorporate in our simulations the dispersive dielectric permittivity of gold \cite{PhysRevB.86.235147}, a complex dielectric constant $n=1.955+0.0045i$, and an effective second-order nonlinear optical coefficient $\chi^{(2)}$ of 6 pm/V for the HfO$_2$-based ferroelectric material embedded in the gap (in the wavelength range considered here the refractive index is almost a constant) \cite{qin2019enhanced}. 

\Cref{fig:1}b (top) shows a typical spectrum of the patch antenna system for normal incidence of TM-polarized plane wave. The system exhibits two resonances (indicated as FP1 and FP3, as we will show later these resonances correspond to Fabry-P\'erot modes of the first- and third-order associated to the gap plasmons) that are ideal candidates to achieve mode-matching for the SHG processes.
In general, however, due to the dispersion of metallic permittivities, it is very difficult for these resonances to satisfy the energy conservation condition, i.e., $\omega_{\rm FP3}=2\omega_{\rm FP1}$.
A rectangular patch, however, allows to overcome this limitation.
By acting on the dimensions of the patch aligned along the $x$ and $y$ directions separately, it is in fact possible to tune almost independently the two resonances.
For example, by increasing the arm-length $a$ (see the dashed-curve in \Cref{fig:1}b), it is possible to largely shift the mode FP1, while only slightly modifying the mode FP3.

For oblique illumination, a distinct resonance indicated by FP2 in \Cref{fig:1}b (bottom) is excited.
The tuning characteristics of this mode is similar to modes FP1 and FP3.
The spatial configuration however is different among all the excited modes, as can be observed from the normalized electric field maps presented in \Cref{fig:1}c$-$e.
The electric field distributions of the modes FP1 and FP3 (for normal and oblique incidence), and FP2 (excited under oblique illumination) can be associated to Fabry-P\'erot modes of the first-, third-, and second-order, respectively.
In what follows, we will realize two optimized designs for SHG, with mode-matching achieved through the interaction of the mode FP1 with either the mode FP2 or FP3.
It is worth mentioning that in general, all geometrical parameters such as thickness of the dielectric spacer, and the lattice constant of the unit-cell contribute to tuning the resonance positions.
Choice of these parameters allows also for controlling the device operating spectral range. 

\begin{figure}
\centering
\includegraphics[width=0.9\linewidth]{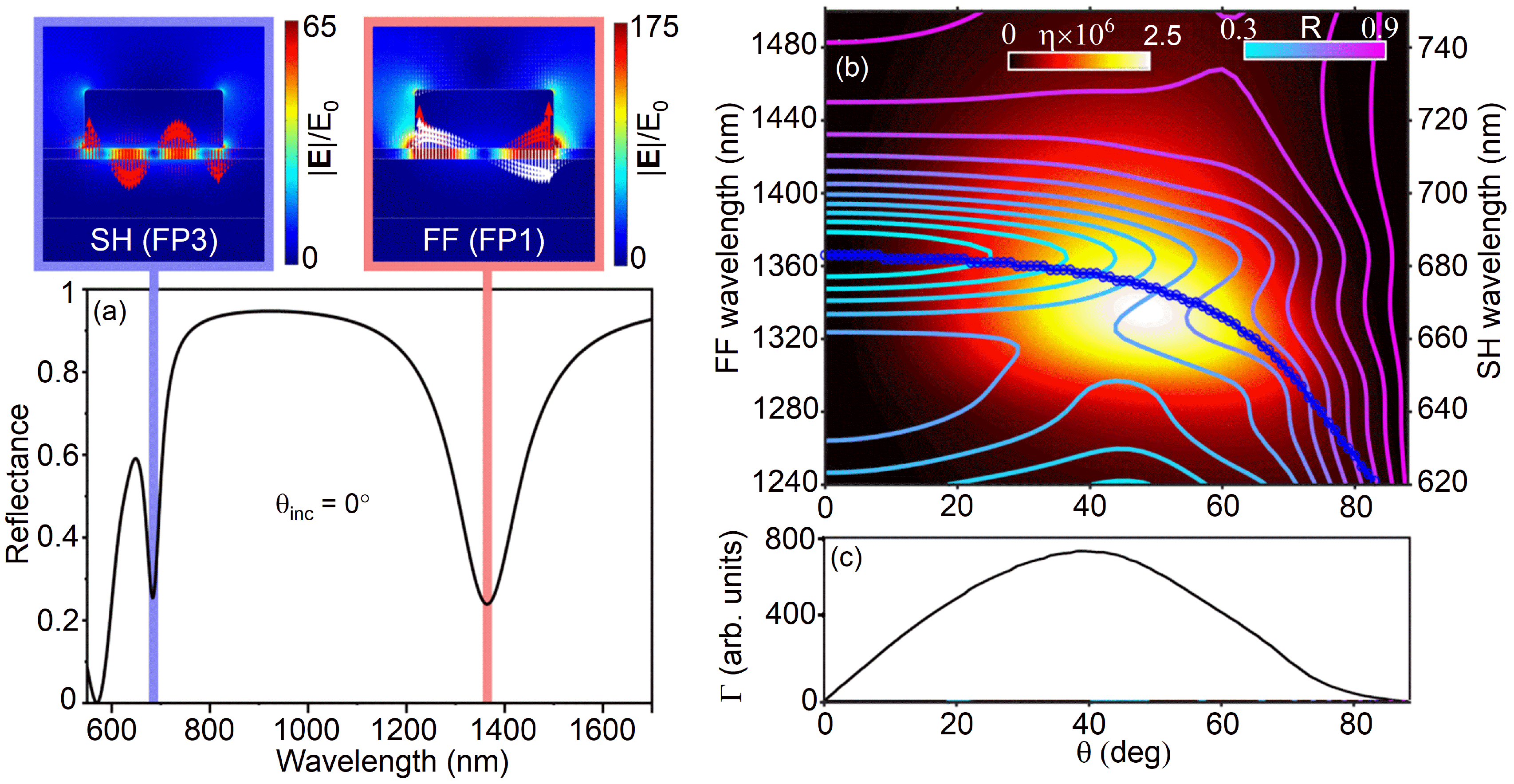}
\caption{The mode-matched design optimized for the interaction of mode FP1 and mode FP3. (a) The linear response: the reflectance spectra and normalized electric field distributions (in the $xz$-plane) of modes at the interacting wavelengths. (b-c) The nonlinear response of the system: (b) the SHG efficiency (heat-map), linear reflection around the SHG wavelength (contour lines), and the FP1 trajectory (blue curve) as a function of the incident angles and wavelengths and (c) the overlap integral extracted following the blue trajectory in (b).}
\label{fig:2}
\end{figure}

In \Cref{fig:2}, we show the electromagnetic response of the mode-matched design optimized for the interaction of modes FP1 and FP3. The design is optimized for a nonlinear conversion of infrared incident to a visible light, as apparent from it's linear reflection spectra presented in \Cref{fig:2} (the optimized design parameters are detailed in the Methods section). The choice of these two modes is driven by three main factors: i) they can be easily excited at normal incidence, ii) the energy matching condition can be realized while keeping reasonable values of the geometrical parameters, and iii) there is a clear overlap between the two modes.
For these reason, one would expect this system to generate the largest SHG conversion, but as we will show there is one more very important factor that one needs to consider. 

In order to numerically evaluate the efficiency of the system, we perform SHG calculations assuming the undepleted pump approximation using a one-way-coupled system of two equations (as detailed in the Methods section).
The conversion efficiency $\eta$ (see Eq. (3) in the Methods section) is evaluated for a TM-incident field carrying an intensity of $I_{\rm FF}\simeq$ 55 MW/cm$^2$, impinging at different angles and wavelengths.
The summary of SHG calculations for the optimized system is shown in \Cref{fig:2}b.
The SHG efficiency map as a function of the incident angle and wavelength is overlapped to the FP1 linear trajectory (blue curve) and linear reflection contour map around the second-harmonic (SH) wavelength.
The system exhibits a modest conversion efficiency of the order of $\eta\simeq1.2\times10^{-9}$ at the normal incidence, where there is a perfect overlap of the modes FP1 and FP3 at the fundamental and second-harmonic wavelengths, respectively (see \Cref{fig:2}b).
For oblique illuminations, however, a gradual increase in the conversion efficiency is observed, with a maximum value reaching three orders of magnitude higher than the efficiency recorded under the normal incidence, $\eta\simeq2.4\times10^{-6}$, at an incidence angle of around incidence $45^{\circ}$.
Note that this maximum in the conversion efficiency does not seem to correspond to any particularly favorable spectral condition. 

In order to understand this behaviour it is useful to introduce the \textit{overlap integral} defined as \cite{Carletti15}:
\begin{equation}\label{eq:1}
\Gamma =\left| \int_\Omega \chi^{(2)} : \mathbf{E}(\omega)\mathbf{E}(\omega)\mathbf{E}(2\omega)^*~d\Omega \right|
\end{equation}
where $\chi^{(2)}$ is the second-order nonlinear susceptibility tensor, while the $\mathbf{E}(\omega)$ and $\mathbf{E}(2\omega)$ are the linear local fields at the fundamental and second-harmonic wavelengths.
The integral is performed over the nonlinear volume $\Omega$.
The overlap integral, $\Gamma$, represents the \textit{propensity} of the energy to flow from the fundamental to the second-harmonic mode.
From its definition, it is clear that in order to maximize $\Gamma$ in \Cref{eq:1}, local field enhancements at the wavelengths of interaction and spatial overlap are not sufficient.
The product of the fields in the integrand needs to add up constructively, i.e., the modes must have the correct symmetry.
It is interesting to remark that for plane waves, $\Gamma$ is maximized when the phase-matching condition ($k_{2\omega}=2k_\omega$), is satisfied.

In order to easily visualize how the fields interact, note that in the film-coupled nanopatch systems, resonant local electric fields in the gap are predominantly polarized perpendicularly to the surface of the metal, i.e., $\mathbf{E}\simeq (0,0,E_z)$.
In \Cref{fig:2}a, the normalized electric field distribution map of the mode excited at the fundamental wavelength, labeled as FF (FP1), reports its $z$-component, $E_z(\omega)$, with white arrows, while red arrows refer to its squared values, $E_z^2(\omega)$. Similarly, the red arrows in the field distribution map of the mode excited at second-harmonic wavelength, indicated as SH (FP3),  correspond to its $z$-component, $E_z(2\omega)$.
It is easy to see how an anti-symmetric mode at the second-harmonic wavelength interacting with the squared of an anti-symmetric mode at the fundamental wavelength (red arrows in the both the field distribution maps) will minimize $\Gamma$, instead of maximizing it.
This can be also observed in the overlap integral shown in \Cref{fig:2}c, calculated along the trajectory of FP1. 
The lowest magnitude of the overlap integral is observed at the normal incidence, which leads to a smaller conversion efficiency despite the field enhancements at the wavelengths of interaction.
For oblique illumination, however, a break in the symmetry results in the increase of the overlap integral, and a gradual increase (peaking in the range of 40-50$^\circ$), as can be observed in the SHG efficiency spectra in \Cref{fig:2}b.

\begin{figure}
\centering
\includegraphics[width=0.9\linewidth]{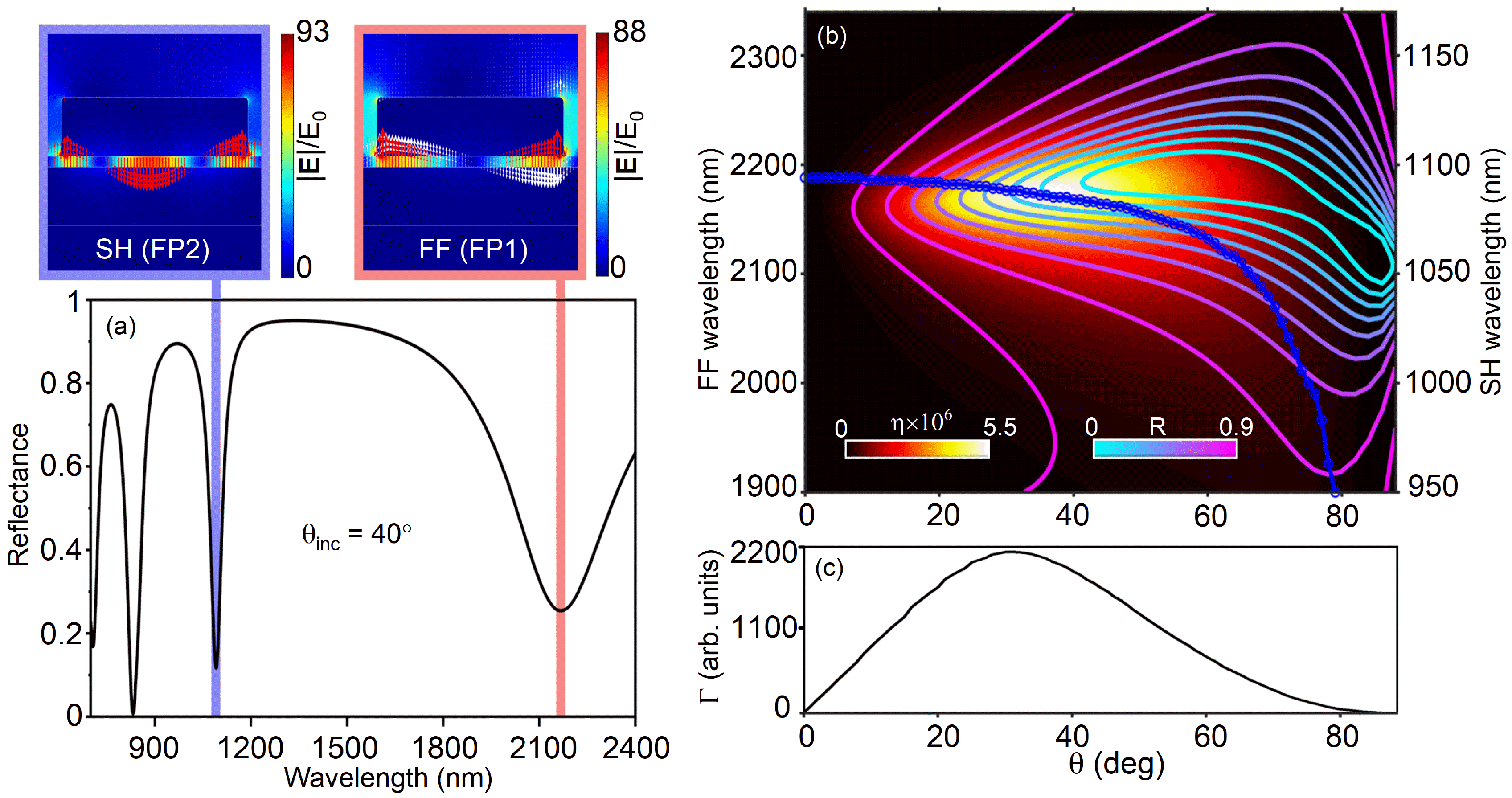}
\caption{The Mode-matched configuration optimized for the interaction of modes FP1 and FP2. (a) The linear reflectance spectra at an incident angle of $\theta=40^\circ$ with the normalized electric field distributions at the resonant wavelengths (in the $xz$-plane). (b-c) The nonlinear response: (b) second-harmonic efficiency spectra (heat-map,linear reflection around the SHG wavelength (contour lines), and the FP1 trajectory (blue curve) as a function of the incident angles and wavelengths and (c) the overlap integral extracted following the blue trajectory in (b).}
\label{fig:3}
\end{figure}

In order to avoid the cancellation effects due to the anti-symmetric nature of the mode FP3, let us now consider the interaction of the mode FP1 with the mode FP2.
The symmetric nature of the mode FP2 should in fact ensure the maximum efficiency conversion. 
The linear characteristics of the mode-matched design optimized for such an interaction are shown in \Cref{fig:3}a, for a TM-polarized excitation impinging at $\theta=40^\circ$. 
This design is optimized to operate in the infrared regime. \Cref{fig:3}b shows the SHG efficiency map as a function of the incident angle and wavelength.
The FP1 linear trajectory (blue curve) and linear reflection contour map around the second-harmonic wavelength are layered on top.
In this case, the lowest values of SHG efficiency at the normal incidence are expected, since the FP2 mode cannot be excited (see the contour levels in \Cref{fig:3}b).
 For oblique illumination, the system exhibits a gradual increase in the SHG efficiency with a peak value of $5.5\times10^{-6}$ at $\theta\simeq 35^{\circ}$.
The peak SHG efficiency for this mode-matched design is increased 2-fold in comparison with the peak efficiency of the previous design.
For completeness we also show in \Cref{fig:3}c the value of the overlap integral $\Gamma$ along the FP1 trajectory in \Cref{fig:3}b.
Similarly to the previous case, $\Gamma$ follows qualitatively the SHG efficiency trend.
Note however that differences between peak angles could be due to the out-coupling efficiency of the mode, which is not considered in the overlap integral calculation.  

\begin{figure}
\centering
\includegraphics[width=0.8\linewidth]{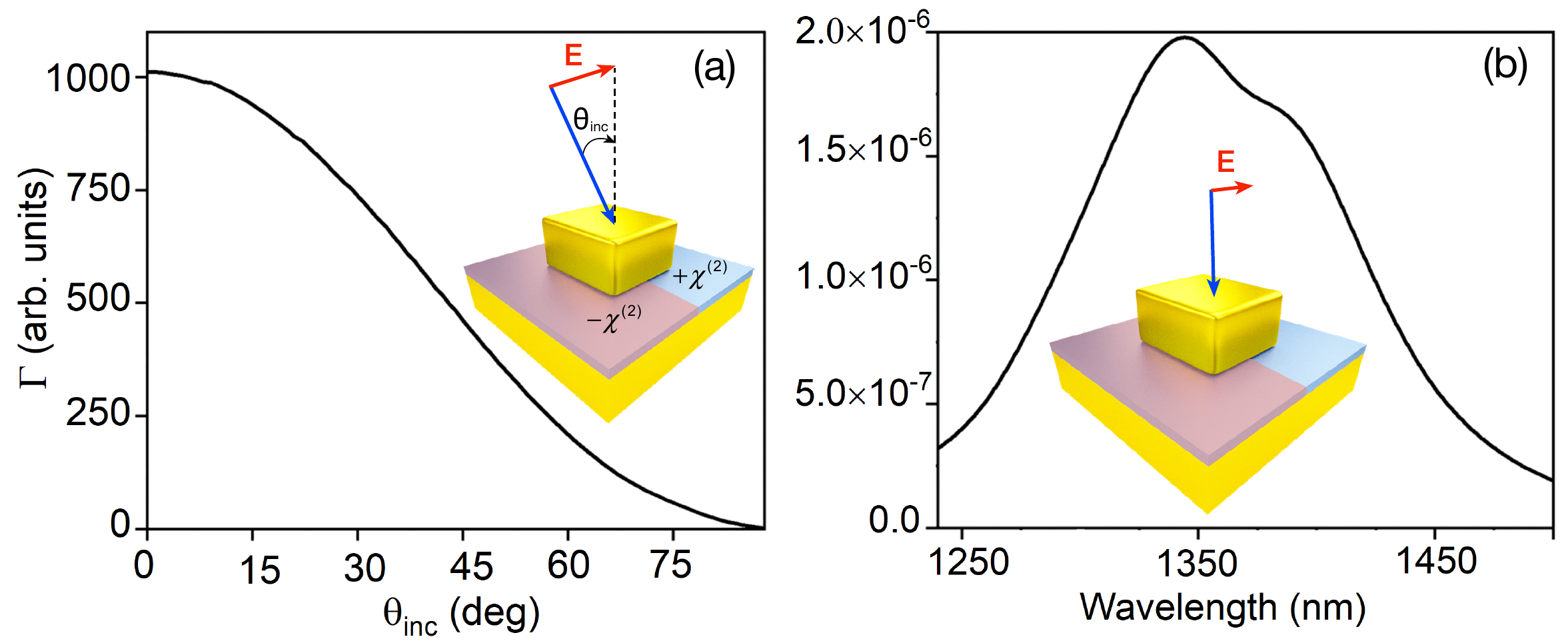}
\caption{(a) The overlap integral along the FP1 trajectory in \Cref{fig:2}b for the poled system depicted in the inset; (b) SHG efficiency spectra at normal incidence.}
\label{fig:4}
\end{figure}

Both configurations analyzed so far require to excite the patch antenna system at an oblique incidence, to achieve maximum possible SHG efficiencies. Ideally, one could remove this inconvenience by using a periodically-poled ferroelectric spacer, such that half of the patch would lay over a $-\chi^{(2)}$ material while the other half on a $+\chi^{(2)}$ material, as shown in the inset of \Cref{fig:4}a. 

In such a configuration, one would be able to optimize the overlap integral between the FP1 and FP3 modes at normal incidence by breaking the symmetry through the sign of $\chi^{(2)}$.
This is shown in \Cref{fig:4}a, where $\Gamma$ is calculated over the FP1 trajectory shown in \Cref{fig:2}b.
This time the maximum magnitude of the overlap integral is obtained at $\theta=0^{\circ}$ where quasi-phase-matching between the modes involved is obtained.
In \Cref{fig:4}b, we show the SHG efficiency for a poled patch antenna system at normal incidence for an interval of frequencies around the modes FP1 and FP3 for the fundamental and harmonic wavelengths, respectively.
As expected, in this case, the conversion is much more efficient ($\eta\simeq 2.0\times10^{-6}$) than the previous case (see \Cref{fig:2}b) at normal incidence.
Interestingly, however, the overall maximum value of $\eta$ remains very close to the maximum value in the previous case.

In conclusion, we have presented a strategy to doubly mode-match plasmonic resonances for efficient SHG using nanopatch antennas in the visible and near-infrared regimes.
We have numerically explored different configurations by providing optimized designs for mode-matching different type of resonances.
Our study has shown that a doubly resonant structure with spatial mode overlap does not guaranteed maximum efficiency.
In fact, destructive interference of the generated fields may lead to weak harmonic conversion.
This can be overcome by exciting the system at a non-zero angle of incidence. 
In general, a measure of propensity of two modes to constructively interact can be obtained by calculating the overlap integral, $\Gamma$, defined in \Cref{eq:1}.
In all our calculations, $\Gamma$, describes very well the trend of the nonlinear SHG efficiency.
Finally, we have shown that efficient SHG at normal incidence can be obtained using a periodically-poled ferroelectric spacer\cite{dicken2008electrooptic,muller2011ferroelectricity}, by correctly aligning poling periodicity with the patch patterning.

The obtained efficiencies are comparable (for analogous input powers) to those obtained with dielectric AlGaAs nanoantennas \cite{Gili:2016hl}, whose nonlinear susceptibility is two orders of magnitude larger than the value considered in this article.
This work show the great potential and versatility of plasmonic nanopatach antennas for nonlinear nanophotonic applications.

\section*{Methods}\label{sec:2}
The SHG process can be described in frequency domain, under undepleted pump approximation, by the following set of equations \cite{boyd2008nonlinear}:
\begin{subequations}\label{eq:2}
\begin{align}
\nabla\times\nabla\times\mathbf{E}_1-k_{1}^2\epsilon(\mathbf{r},\omega)\mathbf{E}_1&=0 \label{subeqn:E2} \\ 
\nabla\times\nabla\times\mathbf{E}_2-k_{2}^2\epsilon(\mathbf{r},2\omega)\mathbf{E}_2&=4\mu_{0}\omega^2\mathbf{P}^{(NL)}\label{subeqn:E3} 
\end{align}
\end{subequations}
where $k_1~=~\omega/c$, $k_2~=~2\omega/c$, with $\omega$ being the fundamental field's angular frequency; $\epsilon(\mathbf{r},\omega)$ is the dispersive permittivity representing the different materials of the design, $\mu_{0}$ is the permeability of free space, and $c$ is the speed of light in vacuum. In the above system of equations, \Cref{subeqn:E2} describes the electric field at the fundamental frequency, $\mathbf{E}_1$, whereas the \Cref{subeqn:E3} is an inhomogeneous vector wave equation that is solved for the generated (second-harmonic) signal $\mathbf{E}_2$.
The right-hand side of \Cref{subeqn:E3} represents the contributions from nonlinear sources in the system.
For simplicity we have considered a nonlinear polarization vector possessing the $z$-component only.
This is justified by the following reasons.
First, the main components of the electric field of the modes excited in the film-coupled nanopatch system, as discussed in the main-text, are primarily polarized perpendicular to the metallic surface (i.e., along the $z$-direction).
Second, the dominant component of the second-order susceptibility tensor exhibited by the nonlinear dielectric material filling the gap between the metallic nanopatch and the film is $\chi_{\perp \perp \perp}^{(2)}$, whose orientation can be controlled through the film growth\cite{muller2011ferroelectricity}.
We defined then the nonlinear polarization vector as $\mathbf{P}^{(NL)}~=(0,0,~\epsilon_0\chi_{\perp \perp \perp}^{(2)}E_{1,z}^2)$, with $\chi_{\perp \perp \perp}^{(2)}=6~$pm/V.
Note that $\mathbf{P}^{(NL)}\ne 0$ only in the thin film embedded in the gap between the metallic film and the nanopatch.

\Cref{subeqn:E2} and \Cref{subeqn:E3} are numerically solved using finite-element based commercial software \textit{COMSOL Multiphysics}, within a customized frequency-dependent implementation.
Solving Eqs. \eqref{eq:2} is a two step process \cite{zeng2009,qin2019enhanced}: first, we solve \Cref{subeqn:E2} under TM-polarized incidence for the fundamental field $\mathbf{E}_1$, in the subsequent step, the second harmonic signal is extracted by solving \Cref{subeqn:E3}, with the nonlinear polarization term, which is defined by utilizing the fields calculated in the first step.
Equations \eqref{eq:2} are solved using periodic boundary conditions, to mimic the electromagnetic response of an infinitely extended periodic array.

The nonlinear conversion efficiency, $\eta$, is defined as \cite{zeng2009}:
\begin{equation}
   \eta~=~\frac{I_{\rm SHG}}{I_{\rm FF}}
\end{equation}
where $I_{\rm FF}$ is incident intensity at the fundamental wavelength $\omega$ and $I_{\rm SHG}$ is the intensity of generated signal measured in the far-field, along the specular direction with respect to the incident excitation.
The intensity of the incident fields considered in the simulations is $I_{\rm FF}\simeq ~55$~MW/cm$^2$.
To circumvent the possible issue of numerical artifacts due to the field localization near the metal corners, we considered rounded corner cube with a radius of curvature of 5~nm.

In the main-text, two optimized mode-matched configurations of the nanopatch system are discussed (see \Cref{fig:2} and \ref{fig:3}). 
Following are the details about the optimized parameters of the unit-cells for both designs.
The unit-cell parameters of the mode-matched design discussed in \Cref{fig:2} are: $a=150$~nm, $b=80$~nm, $g=11$~nm, $h=60$~nm, $t=60$~nm, $A=250$~nm, and $B=200$~nm (refer to schematic of the unit-cell in \Cref{fig:1}a of the main-text). 
Likewise, the configuration discussed in \Cref{fig:3} is realized with the unit-cell parameters: $a=211$~nm, $b=170$~nm, $g=11$~nm, $h=60$~nm, $t=60$~nm, $A=250$~nm, and $B=200$~nm. 

\begin{acknowledgement}
I.-H.L. and S-H.O. acknowledge support from the National Science Foundation (NSF ECCS 1809723 and ECCS 1809240) and Sanford P. Bordeau Endowed Chair at the University of Minnesota. S.A.M. acknowledges the Deutsche Forschungsgemeinschaft, the Engineering and Physical Sciences Research Council (EP/M013812/1) and the Lee-Lucas Chair in Physics. A.R.D. acknowledges support from the Regents of University of Minnesota.
\end{acknowledgement}

\section*{Notes} 
The authors declare no competing financial interest.
\bibliography{biblio}

\end{document}